# Attention Span For Personalisation

Joan Figuerola Hurtado

joan@specifiedby.com

j.figuerolahurtado@napier.ac.uk

## ABSTRACT
A click on an item is arguably the most widely used feature in recommender systems. However, a click is one out of 174 events a browser can trigger. This paper presents a framework to effectively collect and store data from event streams. A set of mining methods is provided to extract user engagement features such as: attention span, scrolling depth and visible impressions. In this work, we present an experiment where recommendations based on attention span drove 340% higher click-through-rate than clicks.

## 1. INTRODUCTION

Feature mining aims to extract features from data streams [1]. These features are then fed into a machine learning (ML) algorithm to complete a certain task. Feature mining is a key problem in ML. The design of those features determines the success of a ML algorithm. Since recommender systems (RS) extensively use ML algorithms, feature mining is also a key problem.

A good feature should be informative, invariant to noise or a given set of transformations, and fast to compute [1]. There are several features for RS such as clicks, ratings, or purchases. A click on an item is arguably the most widely used feature in RS [2,3,4,5]. It is informative in a sense that it shows users' preferences. Furthermore, it is fast to compute, invariant to noise, easy to collect and often non-sparse. However, a click is just one out of 174 events modern browsers can trigger [6]. The data generated by these events is underexplored in feature mining for RS.

Consequently, collecting event-stream data for feature mining becomes a challenging task when there are several events. A server could easily be brought down and a client could become non-responsive if there was an event listener for each event triggered by a web browser. Besides that, 50% of the Internet's traffic is generated by crawlers [10], thus it's important to detect non-human traffic in order to improve data quality and reduce the amount of data to be stored. As a matter of fact, a data collection framework can provide a structured and easy way to capture interactions between entities as well as facilitating the research of the next generation of features for RS.

Some recent research has been focused on attention span. "Attention span is the amount of concentrated time one can spend on a task without becoming distracted. Most educators such as psychologists agree that the ability to focus attention on a task is crucial for the achievement of one's goals." [11] In the domain of webpage ranking, [12,13,14] introduced attention span to improve the relevance of search results and personalise them. So the same search query returns different results depending on the user. YouTube also uses the amount of time a user spent watching a video to make personalised recommendations. The motivation to use attention time was to better surface the videos that viewers actually watch, over those that they click on and then abandon [8].

[7,8] Demonstrated that attention span can slightly improve news recommendations. In that context, attention span refers to the time a user has spent interacting with media content. However, the event collectors presented in [7] do not take into account several scenarios where a user's focus might have faded away, thus giving a false signal of user's attention. Besides that, attention span can not only be applied to news media, it can also be applied to other domains. Attention span could achieve better results by improving the data collectors and the methods to mine it. In addition, visible impressions [9,15] and scrolling depth [16] are also becoming relevant when trying to better understand how users behave online.

This paper provides a novel framework to effectively collect data from event streams. The framework has a low impact on the client and the server. In addition, it is also able to detect and filter out non-human traffic. Furthermore, we present methodology to mine: attention span, scrolling depth and visible impressions. Finally, an experiment is shown where recommendations based on attention span achieved a significant increase in click-through-rates (CTR) compared recommendations based on clicks.

## 2. METHODOLOGY
### 2.1 Data Collection

An event is an action from a user to an item. For example, a user clicks on a product in an ecommerce site. User is an entity, click is an action and product is a target entity. An event is the atomic unit for data collection and its structure is defined as follows:

- *entityId*: identifier of the entity.
- *entityType*: type of the entity.
- *targetEntityId*: identifier of the target entity.
- *targetEntityType*: type of the target entity.
- *type*: event type.
- *timestamp*: time when the event happened.
- *ip*: ip of the client.
- *properties*: properties of the event.

Entity is used instead of user or item. Consequently, the same data structure can be used for user-item, item-item or user-user recommendations.

There are three major methods to emit events: page load, as it happens and pinging. The page load method emits events when a client makes a request to a server to load a page. The main benefit of this approach is that it emits a low number of requests, thus low server workload. However, it does not know when and what a user is doing after the page is loaded.

Secondly, Document Object Model (DOM) events allow JavaScript to register up to 174 different event handlers on elements in a Hypertext Markup Language (HTML) document [7]. Event handlers listen to DOM events and react when triggered. This approach grants real time access to users when a page is loaded. The main drawback of this approach is that events need to be reported to the server as they happen. Consequently, it increases server workload and bandwidth usage, and might affect client responsiveness - particularly on mobile devices.

The third approach is called pinging [17], which consists of reporting events to the server in batches at a certain period of time. In this work, we chose to report every 15 seconds. Hence, each batch stores events that occurred during the last 15 seconds. As a result, pinging resolves the visibility and the workload problem that the previous methods have. A new enhancement is introduced to the pinging method in this paper. The main issue with this method is that it always sends reports to the server regardless whether events occur or not. This unnecessarily increases workload. To solve this issue, events are stored in a list that is emptied after a report is sent. Therefore, if after a 15 second interval the list is empty, the report won't be sent, thus reducing workload.

The client is expected to send large streams of event data to the server. Hence, a rather simple and fast mechanism to store events is required. To avoid blocking the client, reports are sent asynchronously to a queue. Then, workers de-queue events and store them into a database ready to be mined.

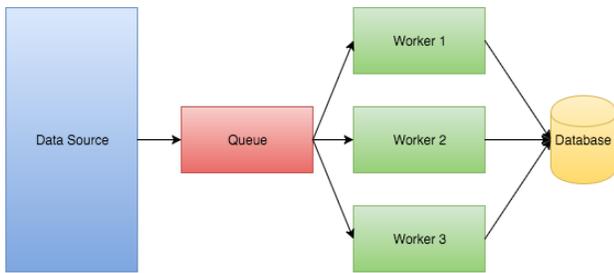

**Figure 1 – Data storage architecture**

Finally, 50% of the Internet's traffic is generated by non-humans [10]. Web crawlers are usually responsible for this traffic. If it is not filtered out, it will affect the quality of the data collected for data mining and the accuracy of the recommendations. To our benefit, the majority of web crawlers do not understand JavaScript [18], meaning that the proposed JavaScript data collectors won't be affected by crawling traffic. However, some search engines such as Google's do understand JavaScript [19]. To prevent this, the data collector script is loaded asynchronously, which is ignored by search engines that do understand JavaScript [20].

## 2.2 Engagement Reports

The engagement report intends to capture engagement signals between user and item. The event structure mentioned before is used to instantiate the engagement report. A simple engagement report can be seen in Appendix 1.

To mine attention span, scrolling depth and visibility, we need to exploit document-object-model (DOM) events. DOM events are triggered when a user interacts with objects in a HTML document. Each event has specific properties that we might want to keep. JavaScript DOM event handlers are registered to the the following events: *Mousemove, Scroll, BeforeUnload, Resize, Focus, DOMContentLoaded, visibilitychange,* and *Keydown*. These are the minimum number of handlers required to mine the features explained in the following sections. A low number of handlers ensures low impact on the client. To reduce the impact further, these handlers are only located on the HTML documents that are items. For instance, a product page in an ecommerce site.

We shall say that a user is engaged with an item when the user triggers one of the DOM events mentioned above. Event handlers share a hash table and when a DOM event is triggered by a user, its handler sets or updates the hash table. The key is the name of the event and the value is an object with its properties. Every five seconds a function checks the hash table. If it's not empty, it's pushed into a list called report. The report stores all the events that were triggered during the last three-five seconds intervals. After that, the hash table is emptied and the same process repeats five seconds later. Following that, an engagement report similar to Appendix 1 is sent, if necessary, to the server every 15 seconds. The length of *report* is between 0 and 3. If it is greater than zero, it means a user engaged with an item. Nevertheless, if the length of the report is zero, it means the handlers did not receive any event so the user did not engage with the item during the last 15 seconds. Therefore, there is no need to send an engagement report.

## 2.3 Feature Mining

Feature mining aims to extract features from data streams [1]. This section will present methodology to mine attention span, scrolling depth and visible impressions.

### 2.3.1 Attention Span

Attention span is the number of seconds a user has spent interacting with a target item. The page load method explained in 2.3 calculates attention span as the time difference between two different page loads. The method provided by [7] calculates attention span by only looking at the following events: *DOMContentLoaded*, *Blur*, *BeforeUnload* and *Focus* loaded. It is more accurate than the page load but overlooks several DOM events that prevent this method from identifying when a user goes idle in many situations. For example, a user has an item page opened and receives a phone call. Or, a user goes to the bathroom leaving the browser with the item page open. These and many more situations create false positives, thus giving an imprecise attention span and directly affecting the performance of the recommender system.

To improve the accuracy of recommendations based on attention span, we need to have an accurate attention span. We assume that if a user triggers one of the DOM events mentioned in the previous section, the user will be engaged with the target item in the next 5 seconds[1]. If not, the user will become idle. Therefore, the total attention time is calculated as follows:

$$totalTime(u,i) = \sum_{1}^{nER} lengthReport_{ui} * 5 \quad (1)$$

, where $u$ is the user, $i$ is the item, $nER$ is the total number engagement reports for user-item and $lengthReport_{ui}$ the number of DOM events during the 15 seconds for that user and item. The latest variable ranges from 0 to 3 as explained in the previous section. Therefore, the maximum time per engagement report is 15 seconds. Note that, a user-item session might have several engagement reports.

### 2.3.2 Scrolling Depth

Scrolling allows users to move vertically in HTML documents, which are often larger than a screen's viewport. Therefore, scrolling is a user engagement signal as it shows that a user wants to know more about the document. For example, in the domain of news outlets, a user who has scrolled 100% of a document shows the user was likely to read the full article. In contrast, a user who is reading an article and scrolls 5% of the document shows the user is probably not as interested in it. This is a particularly relevant feature in domains such as: news outlets, e-commerce, or social media.

---

[1] After running the experiments, we realised that we could improve the accuracy of attention span by analysing the text density in the page layout. For example, if there is low text density and the user triggers an event, we assume the user will be engaged for 5 seconds. If there is high text density and the user triggers an event, we assume the user will be engaged for 10 seconds.

Scrolling events are retrieved from the engagement report in order to mine the scrolling depth. Each scrolling event has four different properties. The first is the maximum distance a user has scrolled from the top of the document. The second is the height of the user's screen, which is likely to differ from users. The third is the current document height, which may vary over time. The fourth is screen width. The following algorithm explains the scrolling depth calculation:

```
function scrolled(screenHeight, maxScrollTop, documentHeight):
   if (screenHeight + maxScrollTop) > documentHeight:
     totalScrolled = documentHeight
   else:
     totalScrolled = screenHeight + maxScrollTop
   return (totalScrolled / documentHeight) * 100
```

$$avgScrolled(u,i) = \frac{\sum_1^{nER} scrolled(sH, maxST, docH)}{nER} \quad (2)$$

, where *u* is user, *i* is item, *sH* is screen height, *maxST* is maximum scroll top and *docH* is document height.

Besides the main content of the document, there are elements that slightly affect this metric such as: header, footer, etc. The accuracy of this metric can be improved by using a CSS identifier or an automated wrapper extractor method [21] to identify the location of the main content. Then, the scrolling handler can use the identifier to register to the main content.

### 2.3.3 Visible Impressions

Scrolling unveils whether an element is within the user's viewport or not. Hence, introducing the viewability metric. It is becoming popular in online advertising because it measures how many ads are truly viewed. According to [9,15], between 42-48% of the ads in the Internet are not viewed. Even though they are counted and paid as such. Therefore, advert viewability optimises the budget of online campaigns and the overall accuracy of advertising platforms.

Viewability is applicable to RS in the following manner. When a user opens an item page, recommendations are often located at the bottom of the page so they are not within the user's viewport. As the user scrolls down, item recommendations show up in the user's viewport. At that precise moment, the impression is stored. On the contrary, traditional methods count impressions when the user opens the page regardless whether a user has seen the recommendations or not.

In addition, viewability can also be applied to other types of pages such as: category pages, search results, social feeds, etc. The main benefit of this metric is the reduction of false positive impressions. Consequently, it improves the accuracy of CTRs as well as increasing it because there are fewer impressions yet the same amount of clicks. The following formula is used to calculate a CTR:

$$CTR(i) = \frac{numClicks}{numVisibleImpression} * 100 \quad (3)$$

,where *CTR* is click-through-rate, *i* is the item, *numClicks* is the total amount of single item views and *numVisibleImpressions* is the total amount of visible impression.

A variation of the engagement report mentioned in section 2.2 is used to collect visible impressions. The scroll handler checks the visibility of items and stores an identifier of the viewed ones into a hash table. A report is sent, if necessary, every 15s seconds with all the viewed items. The report might also be sent before a user leaves the listing page using the *BeforeUnload* handler.

## 3. EVALUATION

Two types of experiments were carried out to evaluate the proposed framework. Both experiments were launched at [Blinded]. The first one consists of evaluating the quality of: attention span, scrolling depth and visible impressions. The second one consists of evaluating recommendations based on attention span against recommendations based on clicks. A demonstration of the features is shown in [22].

### 3.1 Feature Evaluation

In total, users spent 195,745 seconds interacting with 6,041 items[2] with an average attention span per item of 32.4 seconds. In addition, we evaluated the attention span mined by the proposed framework (PF) against the attention span mined by Google Analytics (GA). GA collects data using the page load method described in section 2.1 and PF collects data using the pinging method. A 500-item subset was used for this experiment alongside the event data from PF and GA for one month. GA measured 76,008 seconds and the PF measured 21,805 seconds. GA measured 3.48 times more seconds than PF. Such a big difference makes clear that the page load method is less suitable for measuring attention span. Most importantly, attention span helps to better understand how users interact with items. In a society that constantly requires our attention, it's a valuable asset to know how much attention we got from user.

Moreover, users scrolled 73.32% of the items on average. Looking deeper, a correlation was found between attention span and scrolling as presented in Figure 2. Users who spent more time in a page also scrolled more. The graph contrasts the correlation when the full page was taken into account and when only the main content, excluding: header, footer and other elements. Lastly, users from 15% percentile onwards showed a strong engagement signal so we can infer that they read or checked the whole item.

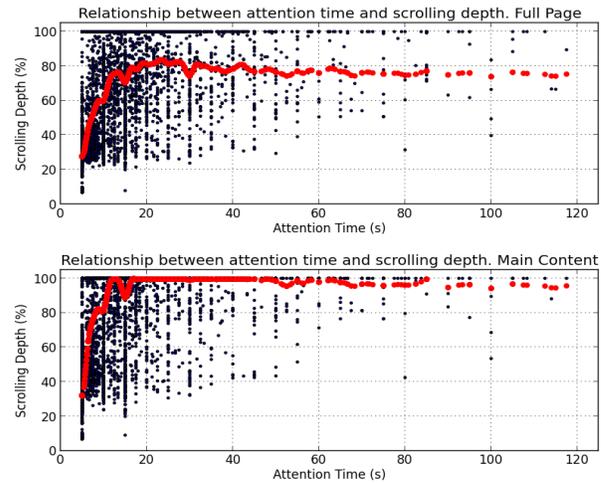

**Figure 2 - Attention span and scrolling depth correlation**

Finally, product and company impressions were collected using the page load method and the ping method for 7 days. The first one counts one impression per item shown in a listing page. In contrast, the second one counts one impression per item viewed. The page load method counted 26.490 impressions and the ping method counted 17.650. That is a 33.37% difference that validates findings from [9,15]. The reduction of false positive impressions increases CTR's by 10.42%. As a result, a more accurate CTR is obtained.

---

[2] To simplify the explanation, we reference products as items.

## 3.2 Recommender System Experiment

The second experiment consists of running an A/B test with two models. One model used average attention span and the other model used clicks as main and only feature. Both models were trained used a vanilla matrix factorization algorithm[3]. In addition, a subset of 2000 items were used for this experiments. Moreover, the experiment ran for a period of time of 2.5 months. Also, the recommendations were shown at the bottom of each item page in a widget with 4 product recommendations. The following table sums-up the results:

**Table 1 – Average Attention Span vs. Clicks**

| Metrics | Avg. Attention Span | Clicks |
|---|---|---|
| Widget Impressions | 506 | 506 |
| Total Clicks | 119 | 35 |
| Widget Avg. CTR | 23.52 | 6.92 |
| Total Impressions | 2024 | 2024 |
| Avg. CTR | 5.88 | 1.73 |

It is clear that the model based on attention span performs substantially better than the model based on clicks. The first one had 3.4 (340%) times more clicks than the second one. Therefore, we validated our hypothesis that attention span is a valuable feature for recommender systems. It is important to remark that we used average attention span rather than total attention span. The latter had equivalent results than the model based on clicks.

## 4. CONCLUSIONS AND FUTURE WORK

In this paper, we presented a framework to effectively collect engagement signals from event-streams. We also proposed enhancements to the pinging technique to reduce server workload and improve client responsiveness. Furthermore, we proposed a methodology to mine attention span, scrolling depth and visible impressions. Finally, we evaluated the mined features with equivalent ones and showed an experiment where attention span substantially increased CTRs in recommendations. Future work will focus on open sourcing the data collection code to foster future research in data collection and feature mining.

---

[3] Note that the purpose of this paper is to provide features that improve any machine learning algorithm - not to improve existing ones.

## 6. APPENDIX

```
{
    'entityId':1,
    'entityType':'user',
    'targetEntityId':10,
    'targetEntityType':'item',
    'ip':'12.345.6.789',
    'timestamp':1459535879,
    'type':'engagement_report'
    'properties':{
        'report':[
            [
                {
                    'scroll':{
                        'document_height':5000,
                        'screen_height':100,
                        'screen_width':980,
                        'scroll_top':300
                    }
                },
                {
                    'mousemove':1
                }
            ], # End first 5 seconds interval
            [
                {
                    'visibilitychange':1
                }
            ], # End second 5 seconds interval
            [
                {
                    'visibilitychange':1
                },
                {
                    'mousemove':1
                },
                {
                    'scroll':{
                        'document_height':5000,
                        'screen_height':100,
                        'screen_width':980,
                        'scroll_top':500
                    }
                },
                {
                    'beforeunload':1
                }
            ] # End third 5 seconds interval
        ]
    }
}
```